\documentstyle[12pt,fullpage,epsfig]{article}
\begin{document}
\begin{center}

\begin{flushright}
     SWAT /262 \\
\end{flushright}
\par \vskip 10mm

\vskip 1.2in

\begin{center}
{\LARGE\bf
Chiral symmetry restoration in the three-dimensional four-fermion model at 
\vskip 0.15in
non-zero temperature and density}

\vskip 0.7in
J.B. Kogut $^a$
and C.G. Strouthos $^{a,b}$ \\
\vskip 0.2in
$^a\,${\it Department of Physics, University of Illinois at Urbana-Champaign,
Urbana, Illinois 61801-3080, U.S.A.}\\
$^b\,${\it Department of Physics, University of Wales Swansea, 
Singleton Park, Swansea, SA2 8PP, U.K.} \\
\end{center}

\vskip 1.0in 
{\large\bf Abstract}
\end{center}
\noindent
The three-dimensional four-fermion model with a $Z_2$ chiral symmetry 
and $N=4$ fermion species is investigated numerically at 
non-zero temparature $T$ and chemical potential $\mu$. The phase diagram 
in the $(\mu, T)$ plane is mapped out  quantitatively. A detailed finite 
size scaling analysis shows that the $T=0$, $\mu \neq 0$ chiral phase 
transition is first order, whereas the $T \neq 0$, $\mu \neq 0$ transition 
remains second order down to very low $T$. Quantitative results for the
location of the end of the first order line are given. The simulations at $T=0$
provided no evidence for a nuclear liquid-gas transition at $\mu < \mu_c$.

\newpage

\section{Introduction}
The chiral phase transition in QCD separating high temperature
(high density) quark-gluon plasma phase from low temperature
(low density) hadronic phase has been studied intensively
in the last decade.
At temperatures of the order of the pion mass, $T \approx 140$ MeV
or densities a few times that of ordinary nuclear matter, $ n \approx (3 - 5)n_0$
with $n_0 \approx 0.15 $ fm$^{-3}$, conditions are reached where hadrons start
overlapping, only the short distance interaction among the internal 
parton degrees of freedom described by QCD is important and dynamical
breaking of chiral symmetry no longer occurs.
Understanding the properties of this transition
is becoming increasingly important in view of recent experimental
efforts to create and detect the quark-gluon plasma in the relativistic
heavy-ion collisions at BNL and CERN.

Since the problem of chiral symmetry breaking and its restoration is
intrinsically non-perturbative, the number of available techniques is
limited and most of our knowledge about the phenomenon comes from lattice
simulations. 
Due to the complexity of QCD, studies have so far
been done on lattices of modest size and have been unable to yield
quantitative results as far as the universality class of the finite temperature
phase transition is concerned. Also, finite density QCD, which is more relevant
to real physics, is a field where lattice
technology is still at an early stage \cite{barbour98}. The complex nature of the 
determinant
of the Dirac operator at finite chemical potential makes it impossible to
use standard simulation algorithms based on positive definite probability functions.

This manuscript addresses the problem of chiral symmetry restoration
at non-zero temperature and non-zero chemical potential in the 
three-dimensional four-fermion model 
\cite{rosen91,kogut93} with a $Z_2$ chiral symmetry
in order to understand what ingredients might play a decisive role in more
complex systems like gauge theories. The model has been simplified as much
as possible in order to produce data of the highest quality and get some insight
into the range of parameters we need for studies of more realistic problems.
The large $N$ description (which is equivalent to a mean field approximation)
of the phase structure of the three-dimensional
four-fermion model at non-zero chemical
potential \cite{hands_kocic_kogut.93} predicts a first order transition for $T=0
$ and a continuous
transition for $T>0$.
The critical value of the chemical potential $\mu_c$ is equal to the
value of the fermion mass at $\mu=0$. Interactions as expected decrease
$\mu_c$ below the mean field result \cite{kim95}.
A recent numerical study of the $T \neq 0, \mu = 0$
transition favors the two-dimensional Ising model exponents for $N<\infty$ \cite{kogut_98}.
In this manuscript we present the results of Monte Carlo simulations  with
a small number of fermion species, i.e., $N=4$ in order to test the
mean field theory predictions and look for possible next-to-the-leading
order corrections.
We show in this paper by employing finite size scaling techniques that the 
$T=0$, $\mu \neq 0$ phase transition is first order. We also show that the second 
order nature of the $T \neq 0$, $\mu=0$ transition remains stable down to 
low temparatures and large values of the chemical potential. A simple 
continuity argument suggests that a tricritical point in the $(\mu, T)$
plane must lie on the low temperature and high density section of the 
critical line. More specifically, our simulations show that the tricritical
point lies on the section of the phase boundary defined 
by: $T/T_c \leq 0.23$, $\mu/\mu_c \geq 0.97$, where $T_c$ is the zero density 
critical temperature and $\mu_c$ is the zero temperature critical chemical
potential. Our simulations did not provide any evidence for non-zero fermion
number density at $T=0$ and $\mu<\mu_c$. This implies that the nuclear liquid-gas
transition is either extremely weak or very close to the chiral transition 
or it does not exist in this model.
Our results show that higher order corrections in the $1/N$ expansion
for the nature and the location of the transition points in the phase diagram are 
very small for this model.

\section{The model}
The continuum spacetime (we work in Euclidean space throughout)
Lagrangian of the model is 
\begin{equation}
\label{lagr1.ff3d}
{\cal L}= \bar{\psi}_i(\partial\hskip -.5em /  
+\mu \gamma_0 +m_q) \psi_i -
\frac{g^{2}}{2 N} (\bar{\psi}_i \psi_i)^{2}.
\end{equation}
Here $\psi_i$ is a four-component spinor and the index $i$ runs over $N$
fermion species. $\mu$ is the chemical potential and $m_q$ is the fermion 
bare mass.
For zero fermion bare mass, ${\cal L}$ has a $Z_2$ chiral symmetry 
\begin{equation} 
\psi \rightarrow \gamma_5 \psi \ ; \ \bar{\psi} 
\rightarrow - \bar{\psi} \gamma_5
\end{equation}
which is spontaneously broken whenever a non-vanishing condensate
$ \langle \bar{\psi} \psi \rangle$ is generated. Both analytical and
numerical work in this model are aided by the introduction of an
auxiliary field $\sigma$, so Eq.~(\ref{lagr1.ff3d}) becomes
\begin{equation}
\label{lagr2.ff3d}
{\cal L}= \bar{\psi}_i(\partial\hskip -.5em / + \sigma)\psi_i
+ \frac{N}{2 g^{2}} \sigma^{2}.
\end{equation}
At tree level, the field $\sigma$ has no dynamics; it is truly an auxiliary 
field. However, it acquires dynamical content by dint of quantum effects 
arising from integrating out the fermions. In the chiral limit the 
vacuum expectation value of $\sigma$ becomes a dynamical fermion mass
and it is a convenient order parameter for any chiral phase transition 
in the theory.

There are several motivations for studying such a simple model: (i) for sufficiently 
strong
coupling it exhibits chiral symmetry breaking at zero temperature and density;
(ii) the spectrum of excitations contains both baryons and mesons, i.e. the elementary
fermions and the composite fermion$-$anti-fermion states; (iii) the model has an interacting
continuum limit; (iv) when formulated on the lattice it has a real Euclidean action
even for non-zero chemical potential; (v) the Yukawa-like coupling makes the 
inverse Dirac operator non-singular
and the simulations can be done in the chiral limit directly, whereas in gauge theories
the matrix to be inverted becomes ill-conditioned and the simulations have to be done
using a fermion bare mass; (vi) the model has fewer degrees of freedom than QCD, 
and can be studied with greater precision on much bigger lattices than the lattices presently 
used for QCD thermodynamics.
The chiral symmetry must be discrete in order to observe a
phase transition at non-zero temperature in $d=2+1$, otherwise according to 
Coleman$-$Mermin$-$Wagner theorem \cite{coleman}
a continuous chiral symmetry must be manifested algebraically for all $T>0$.

In four dimensions the four-fermion model is a trivial theory \cite{hands98} and is 
believed to be an effective theory of quarks and gluons at 
intermediate energies. Unfortunately, the order of the chiral phase transition 
at non-zero chemical potential in this model depends on the value
of the cutoff \cite{klevansky}. In two dimensions the discrete chiral
symmetry restoration is dominated by the materialization of kink$-$anti-kink
states. 
The zero temperature and finite chemical potential transition is 
first order, whereas 
for any finite number of fermion species $N$ the chiral symmetry
is restored at any non-zero temperature due to the condensation of the 
kinks \cite{karsch}.

%\section{Lattice formulation}
The four-fermion model in its bosonized form may be formulated on a 
spacetime lattice using the following action:

\begin{equation}
S_{lat}=\sum_{i=1}^{N/2} \left( \sum_{x,y} \bar{\chi}_{i}(x) M_{x,y} \chi_{i}(y)
+\frac{1}{8} \sum_{x} \bar{\chi}_{i}(x) \chi_{i}(x) \sum_{ \langle \tilde{x},x \rangle}
\sigma(\tilde{x}) \right) +\frac{N}{4g^{2}} \sum_{\tilde{x}} \sigma^{2}(\tilde{x}),
\label{gnaction_lattice}
\end{equation}
where $\chi_{i}$ and $\bar{\chi}_{i}$ are Grassmann-valued staggered fermion fields
defined on the lattice sites, the auxiliary field $\sigma$ is defined on the dual lattice
sites \cite{cohen}, and the symbol $\langle \tilde{x},x \rangle$ denotes the set of 8 dual lattice
sites $\tilde{x}$ surrounding the direct lattice site $x$. The lattice spacing $a$ has been
 set to one for convenience. Details and motivation of this particular scheme can be found in
\cite{kogut93}. The fermion kinetic operator $ M $ is given by
\begin{equation}
M_{x,y} = \frac{1}{2} \left[ e^{\mu} \delta_{y,x+\hat{0}} - e^{-\mu} \delta_{y,x-\hat{0}} \right]
+ \frac{1}{2} \sum_{\nu=1,2} \eta_{\nu}(x) \left[ \delta_{y,x+\hat{\nu}} -
\delta_{y,x-\hat{\nu}} \right],
\end{equation}
where $\eta_{\nu}(x)$ are the Kawamoto-Smit phases $(-1)^{x_0+...+x_{\nu-1}}$. 
The influence of the chemical potential $\mu$ is manifested through the 
timelinks following \cite{chem.latt}. Only fermion loops which wrap 
around the timelike direction are affected by its inclusion.
The energy density $\epsilon$, and the fermion number density
$n$, are defined by
\begin{equation}
E = - \frac{1}{V_{s}} \frac{ \partial \ln Z}{\partial \beta}
= \frac{1}{V} \mbox{tr} \partial_{o} \gamma_{o} S_{F} = \frac{1}{2 V}
\langle \sum_{x} e^{\mu} M^{-1}_{x,x+\hat{0}} - 
e^{-\mu} M^{-1}_{x,x-\hat{0}} \rangle,
\end{equation}
\begin{equation}
n = - \frac{1}{V_{s} \beta} \frac {\partial \ln Z}{\partial \mu}
= \frac{1}{V} \mbox{tr}\gamma_{o} S_{F} = \frac{1}{2V}
\langle \sum_{x} e^{\mu} M^{-1}_{x,x+\hat{0}} + 
e^{-\mu} M^{-1}_{x,x-\hat{0}} \rangle.
\end{equation}
Here $V_s$ is the spatial volume and $V=V_s \beta$ the overall volume of 
spacetime.
The final expression in each case is the quantity measured in the simulation,
using a noisy estimator to calculate the matrix inverses.                     

The cubic lattice has
$L_{s}$ lattice spacings $a$ in spatial directions and $L_t$ lattice
spacings in the temporal direction.
The momentum cutoff scale can be defined as $\Lambda=
1/a$, the temperature is given by $T=1/(L_ta)$ and the mass of the 
sigma meson is $m=1/(\xi a)$,
where $\xi$ is the correlation length of the sigma field.
To reach a continuum limit one has to satisfy the following two
conditions: $\Lambda\gg T$ and $\Lambda\gg m$. The condition $\Lambda\gg T$
requires a lattice with sufficiently large $L_t\gg 1$. The parameters
of the action should then be tuned towards their critical values
where the correlation length $\xi \gg 1$. This satisfies the
condition $\Lambda\gg m$.

\section{Monte Carlo simulations}

We performed Monte Carlo simulations of the four-fermion model in
$d=2+1$ dimensions to study its phase structure in the $(\mu, T)$ plane.
We used a hybrid
Monte Carlo method described in \cite{kogut93}, which proved to be
very efficient for our purposes.  Since the chiral symmetry is
discrete we were able to simulate the model directly in the chiral
limit, i.e., setting the fermion bare mass to zero. 
The three-dimensional character of the model and its relatively simple
form allowed us to perform simulations on large lattices and generate
a large number of trajectories relative to state-of-the-art lattice
QCD simulations.
This allowed  a particularly accurate determination of the
order of the phase transition at various points in the $(\mu, T)$ plane.
In order to check whether it is possible to detect next-to-the-leading 
order corrections to the large $N$ calculation we set $N=4$. 
This is the smallest number of fermion species allowed by the 
hybrid Monte Carlo algorithm, because in order to have a positive
semi-definite fermionic determinant, $N$ must be a multiple of four.

We used the following two methods to optimize the performance of the
hybrid Monte Carlo procedure. The first method consisted of tuning the
effective number of fermion flavors $N^{'}$, which is used
during the integration of the
equations of motion along a microcanonical trajectory, so as to
maximize the acceptance rate of the Monte Carlo procedure for a fixed
microcanonical time-step $d\tau$. As the lattice size was increased,
the time step $d\tau$ had to be taken smaller and the optimal $N'$
approached $N$.  For example, for the  $N=4$ theory on a $16
\times 32^{2}$ lattice the choices $d\tau=0.12$ and $N'=4.026$
gave acceptance rates greater than $90\%$ for all couplings of
interest. To maintain this acceptance rate on a $16 \times 64^{2}$
lattice we used $d\tau=0.10$ and $N'=4.018$.  
The Monte Carlo procedure was also optimized by choosing the
trajectory length $\tau$ at random from a Poisson distribution with
the mean equal to $\bar{\tau}$. This method of optimization, which
guarantees ergodicity, was found to decrease autocorrelation times
dramatically \cite{HaKoKoRe94}. For most of our runs we used the
average trajectory length $\bar{\tau} \simeq 2.0$.  As usual, the
errors were calculated by the jackknife
blocking, which accounts for correlations in a raw data set.

\subsection{$\mu = 0$, $T=0$}
\label{m=0.t=0}
\begin{figure}[p]

                \centerline{ \epsfysize=2.8in
                             \epsfbox{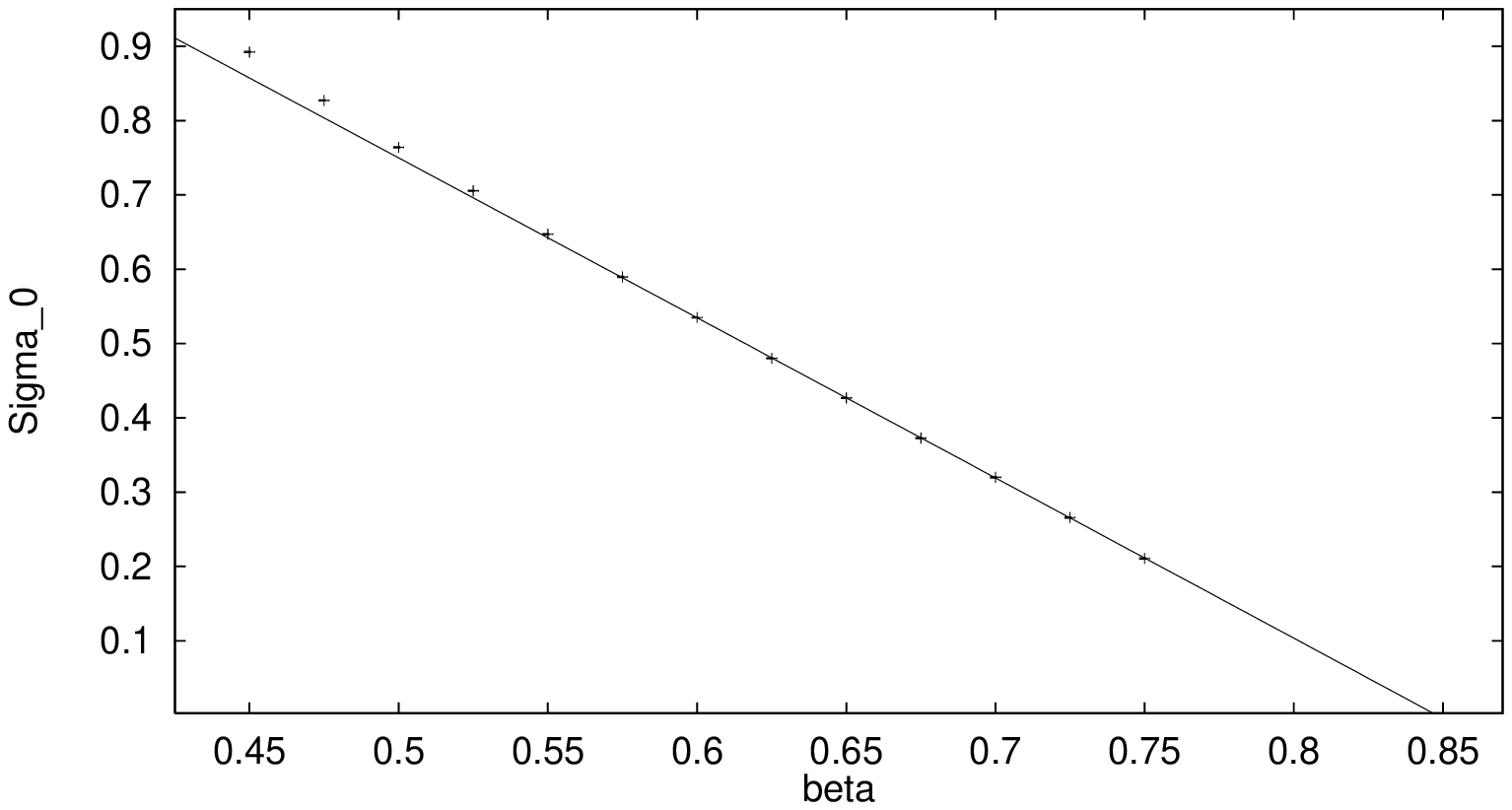}}

\smallskip
\caption[]{Order parameter $\Sigma_0$ vs. $\beta$ at $T=\mu=0$.}
\label{fig:sigma_t=0_mu=0}
\end{figure}

In this section we present the results of the simulations at 
zero temperature and zero chemical potential. The purpose 
of these simulations is to determine the scaling window of the theory at
$T=0$ and $\mu=0$. 
The knowledge of the scaling window is necessary in order 
to ensure that the value of the 
lattice coupling $\beta \equiv 1/g^2$ used in the simulations at 
non-zero temperature
and density is sufficiently close to the bulk
critical coupling $\beta_c$, which means that lattice artifacts are negligible.
In other words, we verified that another important physical parameter,
the zero-temperature mass, $m_0$, is sufficiently smaller than the cutoff
$\Lambda$. We ran on large symmetric lattices, such as $20^3$ with
$N=4$ for $\beta=[0.500, 0.750]$
and determined the magnetic critical exponent $\beta_{mag}$
by fitting the values of the order parameter $\Sigma_0 \equiv \langle \sigma \rangle$ to the 
scaling relation $\Sigma_0 = A(\beta_c - \beta)^{\beta_{mag}}$. 
We found that $\beta_c \approx 0.845$ and 
$\beta_{mag}=1.00(2)$ which is in agreement with the 
analytical prediction
$\beta_{mag}=1+{\cal O}(1/N^2)$ for the $T=0$, $\mu=0$ scaling \cite{kogut93}. 
According to the graph in Fig.~\ref{fig:sigma_t=0_mu=0}, the range of $\beta$ 
for which the $(\Sigma_0, \beta)$ points fall on the `scaling' line is 
$\beta \approx [0.55,0.75]$. 
This confirms that 
for these values of the coupling the lattice theory remains in the
scaling window for all range of temperatures down to $T=0$ and effects of the
lattice are negligible. 

\subsection{$\mu \neq 0$, $T = 0$}
\label{m.neq.0_t=0}
In this section we discuss the results from numerical simulations of the 
model at non-zero chemical potential and zero temperature.
The mean field theory
approximation \cite{hands_kocic_kogut.93} predicts that the finite $\mu$ phase transition
is everywhere second order except at the isolated point $T=0$. 
The issue raised by our lattice simulations in this section is
whether the first order nature of the transition predicted by mean field theory
at $T=0$ persists, when the theory has a small number of fermion species.
In addition these simulations are motivated by our interest to study the difficulties 
in determining the order
of the transition using detailed finite size scaling methods.
We simulated the system at $\beta=0.625$ because this value of $\beta$ is deep 
in the broken phase and is also 
in the scaling window of the bulk critical point which means we can extract 
continuum physics from these simulations by using standard methods. The order parameter
for these simulations is defined by $\Sigma \equiv \langle |\sigma| \rangle$. The 
absolute value of $\sigma$ is necessary on a finite size lattice 
in order to take into account the tunneling events between 
the degenerate $Z_2$-symmetric vacua when the parameters are tuned close to the transition.  
In the thermodynamic limit $\Sigma \rightarrow \langle \sigma \rangle$.

\begin{figure}[p]

                \centerline{ \epsfysize=3.0in
                             \epsfbox{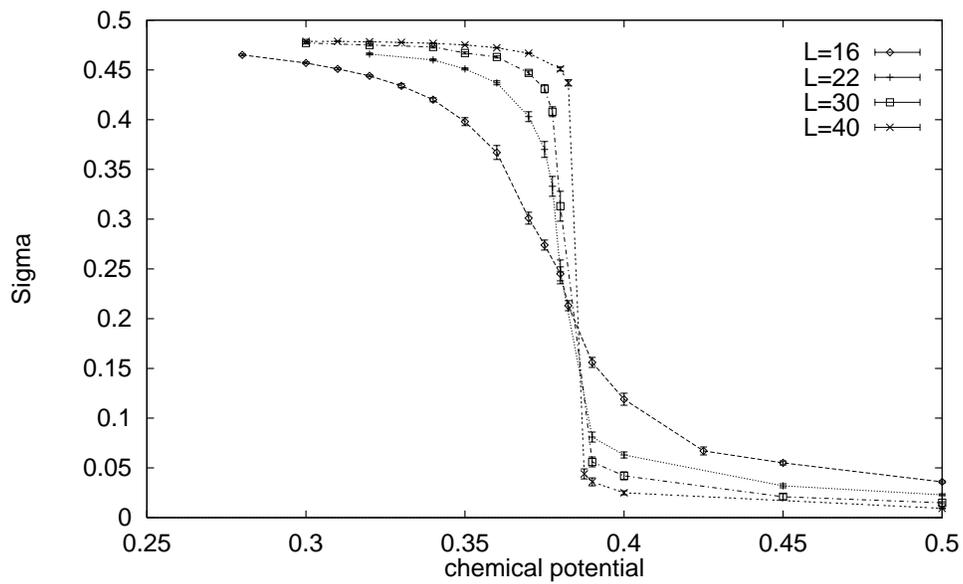}}

\smallskip

\caption[]{Order parameter $\Sigma$ vs. $\mu$ at $T=0$
for $L=16, 22, 30, 40$.} 
\label{fig:sigma_t=0_mu.nonzero}
\end{figure}

\begin{figure}[]

                \centerline{ \epsfysize=3.0in
                             \epsfbox{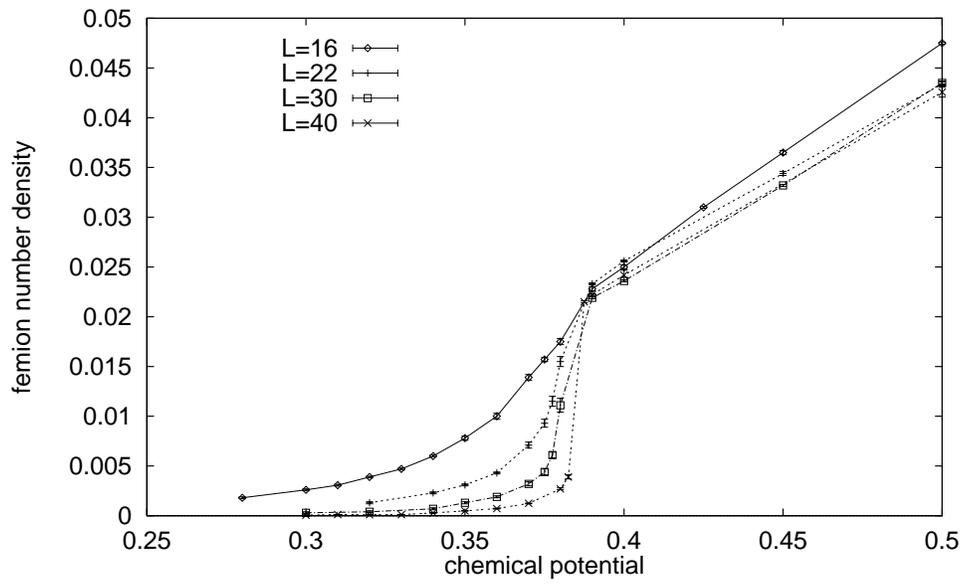}}

\smallskip
\caption[]{Fermion number density $n$ vs. $\mu$ at $T=0$ for
$L=16, 22, 30, 40$.}
\label{fig:density_t=0_mu.nonzero}
\end{figure}

Figure~\ref{fig:sigma_t=0_mu.nonzero} shows the order parameter $\Sigma$ versus $\mu$ 
for fixed $\beta=0.625$
on $16^{3}$, $22^{3}$, $30^{3}$ and $40^{3}$ lattices.
The values of $\Sigma$ on the bigger lattices show no $\mu$ dependence
until $\mu$ reaches the vicinity of the critical chemical potential $\mu_c$.
The variation of $\Sigma$ becomes more abrupt as the volume of the system
increases, indicating that a step-function type singularity will develop
in the thermodynamic limit.
The data on the $40{^3}$ lattice
show a jump discontinuity in $\Sigma$
as $\mu$ varies from $0.3850$ to $0.3875$.
The fermion number density $n$ is 
plotted in Fig.~\ref{fig:density_t=0_mu.nonzero} 
as a function of $\mu$ for various lattice sizes. The transition at 
$\mu_c \approx 0.3850$ is seen equally well in this quantity. Futhermore, the variation
of $n$ near the transition becomes more abrupt as the volume increases,
indicating again that a step-function singularity will develop in the thermodynamic limit.
It is clear from the behavior of the various observables that relatively large lattices
are required to simulate the theory's actual critical behavior. The discontinuous
transition seen on a $40^{3}$ lattice is replaced by a relatively smooth curve
on the $16^{3}$ lattice.

In order to compare the fermion dynamical mass $m_f$ at $\mu=0$ with with the critical 
value of the chemical potential $\mu_c$
we calculated the fermion propagator $G({\bf x}, t)$ on the $40^{3}$ lattice.
Then, we formed a zero momentum fermion propagator
\begin{equation}
C_f(t)=\sum_{\bf x} G({\bf x}, t).
\end{equation}
After calculating the average and the covariance matrix of $C_f(t)$ we fitted it
to the following functional form:
\begin{equation}
C_f(t)=A[e^{m_f t}-(-1)^{t}e^{-m_f(L_t-t)}].
\end{equation}
The value of the fermion dynamical mass is $m_f=0.403(1)$, which is according to our 
expectation very close 
to $\mu_c$. The small discrepancy is due to interaction and finite size effects.

\begin{figure}
  \begin{center}
    \begin{tabular}{c}
%      (a)  $L=16, 22$\\
      \centerline{ \epsfysize= 3.0in
                   \epsfbox{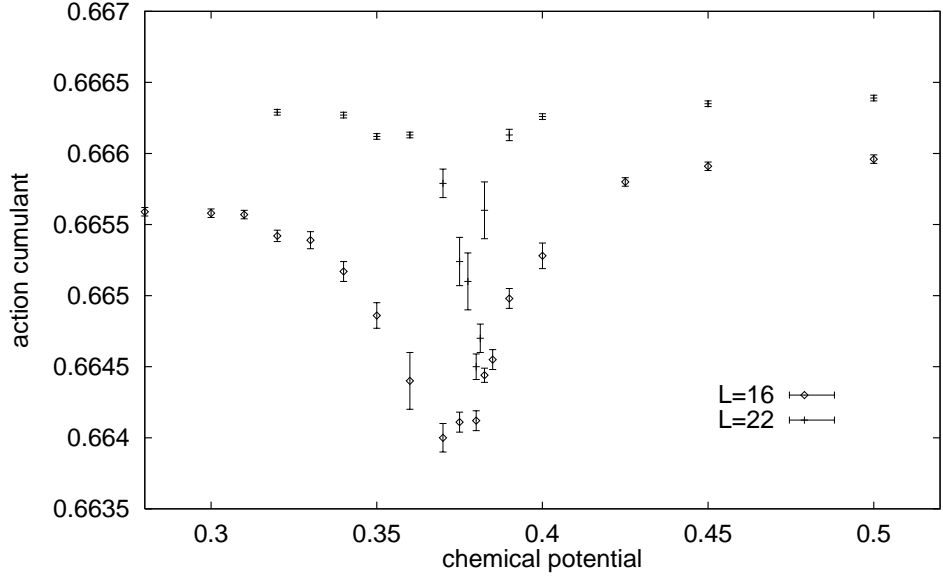}} \\ \\ \\
%      (b) $L=18, 30$ \\
       \centerline{ \epsfysize= 3.0in
                   \epsfbox{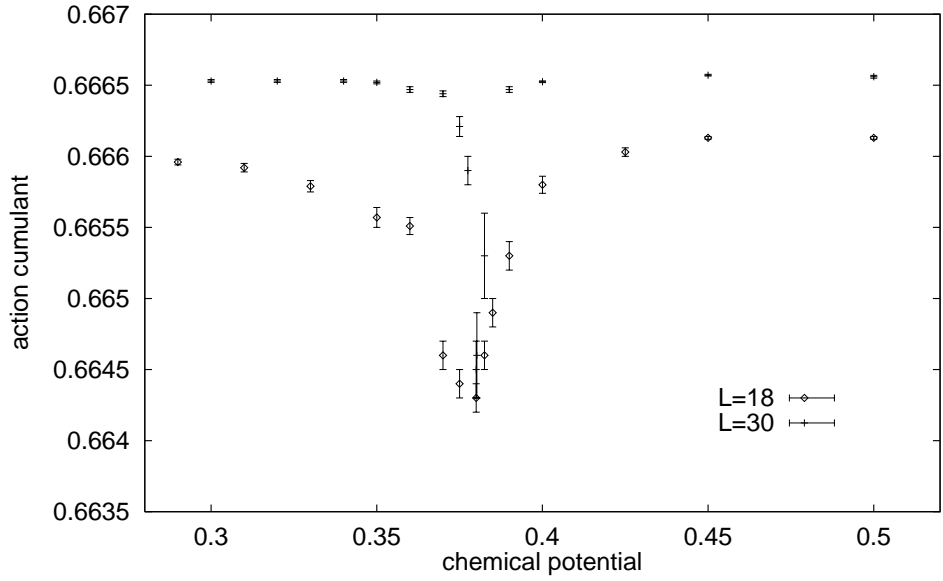}}
    \end{tabular}
  \end{center}
  \caption{The bosonic action cumulant $B_S$ vs. $\mu$ at $T=0$
           for $L=16, 18, 22, 30$.}
\label{fig:cumulant_t=0_mu.nonzero}
\end{figure}

\begin{figure}[]

                \centerline{ \epsfysize=3.0in
                             \epsfbox{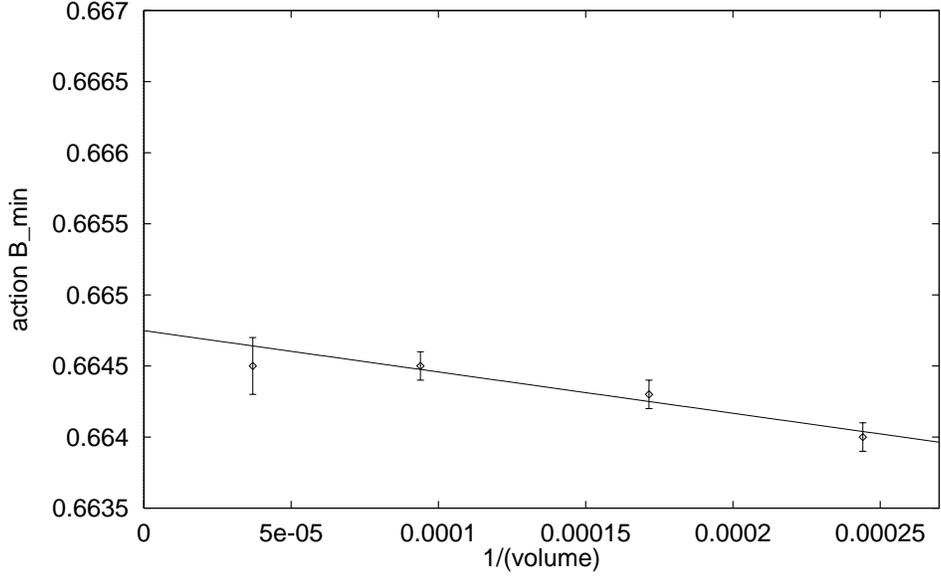}}

\smallskip
\caption[]{The minimum values of the bosonic action cumulant $B_S$ vs. inverse volume
           at $T=0$.
          The solid line represents the best linear fit to the data.} 
\label{fig:cumulant_min.t=0_mu.nonzero}
\end{figure}

Another interesting indicator which is often used to determine the order of the 
transition was proposed
by Binder and Landau \cite{binder_84}, namely the reduced cumulant of the bosonic action $S$
defined by 
\begin{equation}
B_S=1- \frac{\langle S^{4} \rangle}{3 \langle S^{2} \rangle^2}.
\end{equation}
This quantity should approach $\frac{2}{3}$, if the distribution of the bosonic 
action is described 
by a single Gaussian form with the width vanishing at infinite volume, such as in the case of
a second order transition; otherwise it deviates from this value. 
In practice we calculate 
$B_S$ for several values of the chemical potential near the transition
for each of several lattice sizes. Then a plot of $B_S$ versus chemical potential
is made for each lattice size. An accurate value of $B_{S,min}$ for each lattice size
is determined and then $B_{S,min}$ is plotted versus $\frac{1}{V}$. If the transition
is second order, this plot should have a $y$-intercept (representing the infinite volume
limit) approaching the value of two-thirds.
In order to locate $B_{S,min}$ accurately we performed simulations at very small intervals
of $\mu$ near the transition. 
We accumulated approximately 30,000 to 40,000 trajectories for the various 
lattice sizes at each value 
of $\mu$ near $\mu_c$.
The cumulant as a function of $\mu$ is plotted in 
Fig.~\ref{fig:cumulant_t=0_mu.nonzero} for $L=16,18,22,30$. 
In Fig.~\ref{fig:cumulant_min.t=0_mu.nonzero} we plot $B_{S,min}$ versus inverse volume
$\frac{1}{V}$. 
The straight line in Fig.~\ref{fig:cumulant_min.t=0_mu.nonzero} represents
the best linear fit to the data and its $y$-intercept is equal to  0.66475(14), 
which is distinctly different from the value of $\frac{2}{3}$, indicating
clearly that the 
model has a first order finite density and zero temperature phase transition. 
A similar result was obtained from the analysis of the reduced cumulant 
behavior of the energy density.

Our simulations were unable to distinguish a possible nuclear matter
liquid-gas transition from the chiral transition. In the large $N$ approximation 
the two transitions coincide \cite{hands_kocic_kogut.93}.  The data in
Fig.~\ref{fig:density_t=0_mu.nonzero} for $n$ versus $\mu$
show only one phase transition at $\mu_c$. The smooth tail in the density
before the transition is a finite size effect which dies out as the
volume of the system increases.
We extended our search for a non-zero $n$ at $\mu<\mu_c$ by performing
new simulations on large lattices ($30^3, 40^3, 54^3$) at $\beta=0.45$ which is far away from
the continuum limit. In this case the correlation length is smaller than before which
implies smaller finite size (temperature) effects. The chiral phase transition at this value
of the coupling takes place at $\mu_c=0.692(1)$. Again, we did not detect a size-independent
saturation in the fermion number density at $\mu<\mu_c$. 
At $\mu=0.690$ the fermion number density $n$ is $0.0118(4)$ on the $30^3$ lattice,
$0.0040(2)$ on the $40^3$ lattice and $0.00126(5)$ on the $54^3$ lattice.
This implies that for $N=4$ the next-to-the-leading
order corrections are very small if not zero, in the sense that either
the two transitions are very close or the nuclear matter liquid-gas
transition is so weak that is smoothed out by finite size effects or it doesn't
exist in this model.

\subsection{$\mu \neq 0$, $T \neq 0$}
\label{m.neq.0_t.neq.0}
In this section we present the results from the simulations of the 
four-fermion model at non-zero temperature and chemical potential.
The purpose of these simulations is to study the phase structure
of the model in the $(\mu, T)$ plane.
The large $N$ result predicts a second order phase transition for $T>0$.
We varied $T$ by changing either the lattice temporal 
extend $L_t$
or the lattice spacing $a$ which according to the discussion in Sec.~\ref{m=0.t=0} 
vanishes when the coupling $\beta$ 
is tuned to its 
$T=\mu=0$ critical value $\beta_c \approx 0.845$.                     

\begin{figure}[htb]

                \centerline{ \epsfysize=3.0in
                             \epsfbox{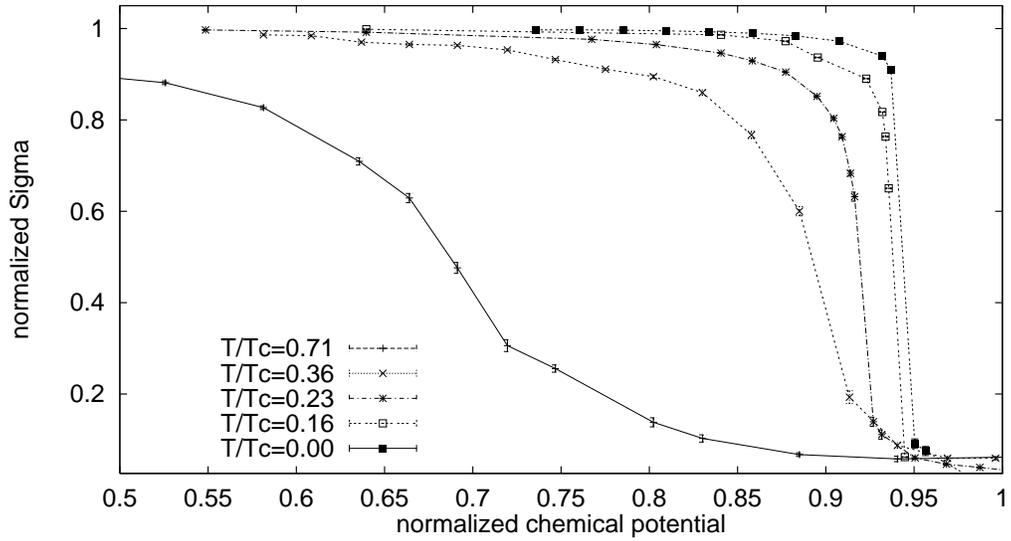}}

\smallskip

\caption[]{$\Sigma(T,\mu)/\Sigma(T=0,\mu)$ vs. $\mu/m_f$ for different values 
of temperature $T$.}
\label{fig:sigma_t_mu_normalized}
\end{figure}

\begin{figure}[htb]

                \centerline{ \epsfysize=3.0in
                             \epsfbox{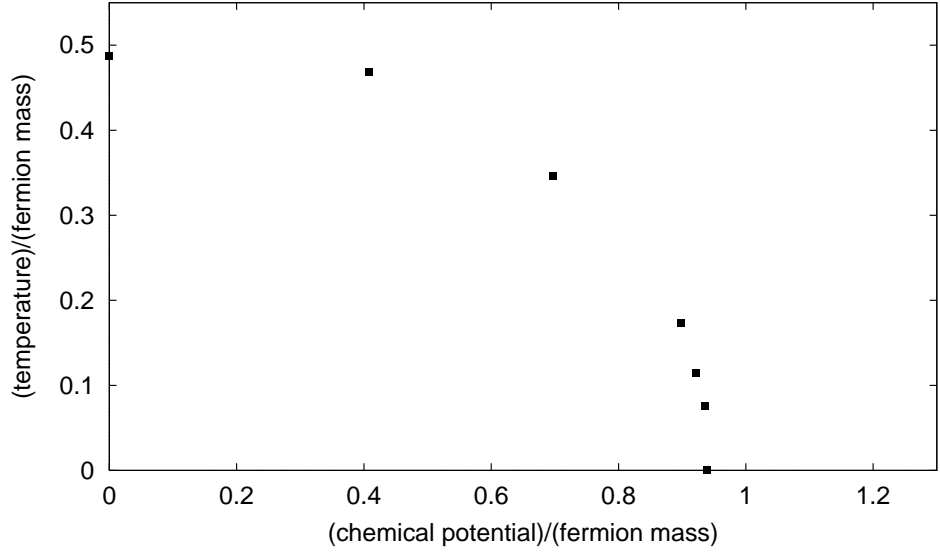}}

\smallskip
\caption[]{Phase diagram in the $(T/m_f, \mu/m_f)$ plane. }
\label{fig:phase.dgr}
\end{figure}

In Fig.~\ref{fig:sigma_t_mu_normalized} we plot the normalized 
order parameter $\Sigma(T,\mu)/\Sigma(T=0,\mu)$ as a function of $\mu/m_f$ at different values
of $T$.  This normalization is essential
in order to convert the order parameter and the chemical potential 
into physical units by getting 
rid of the lattice spacing. 
It therefore enables us to compare the behavior of the 
order parameter as a function of the chemical potential at different 
temperatures on lattices with different lattice spacings.
According to the discussion in Sec.~\ref{m=0.t=0} the lattice discretization effects
cannot be neglected when $\beta$ is smaller than $0.55$.
The parameters for the curves in Fig.~\ref{fig:sigma_t_mu_normalized} 
from left to right in terms of 
$(\beta$, lattice size, $T/T_c)$ are: 
$(0.65, 8\times 48^{2}, 0.71)$, $(0.65, 16\times48^{2}, 0.36)$, $(0.55, 16\times48^{2}, 0.23)$,
$(0.55, 24\times48^{2}, 0.16)$, $(0.625, 40^{3}, 0.0)$.   
It is clear from the shapes of these curves that the transition becomes sharper
as we decrease the temperature.
In Fig.~\ref{fig:phase.dgr} we map out the phase diagram of the model 
in the $(\mu/m_f, T/m_f)$ plane. As the temperature increases from zero,
the chiral condensate `melts' at a smaller value of the  
critical chemical potential $\mu_c(T)$ 
which suppresses energetically the quark$-$anti-quark pairing.

After having determined the order of the phase transition
at zero temperature (see previous section), we extended our work to determine
the order
of the transition at the three points of Fig.~\ref{fig:phase.dgr} 
with the lowest non-zero values of $T/T_c$. 
At these points the variation of $\Sigma(T,\mu)/\Sigma(T=0, \mu)$ as a function of 
$\mu/m_f$ is sharp near the transition.

\begin{figure}[htb]

                \centerline{ \epsfysize=3.0in
                             \epsfbox{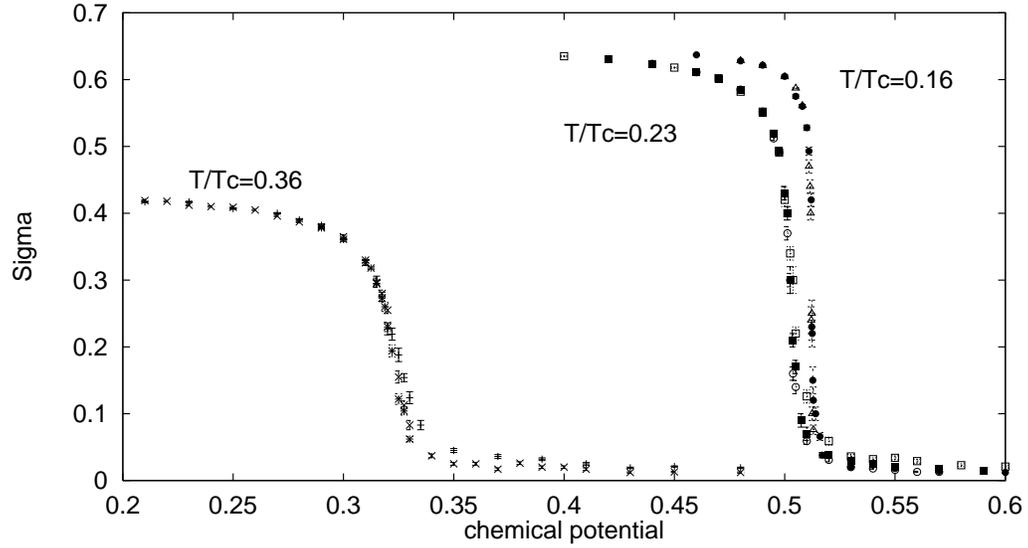}}

\smallskip
\caption[]{$\Sigma$ vs. $\mu$ for $T/T_c=0.36$, $(L_s=32, 48, 64)$; $T/T_c=0.23$, 
$(Ls=32, 48, 64)$ and $T/T_c=0.16$, $(L_s=48, 72)$.}
\label{fig:sigma_mu.non0_t.non0}
\end{figure}

\begin{figure}[htb]

                \centerline{ \epsfysize=3.0in
                             \epsfbox{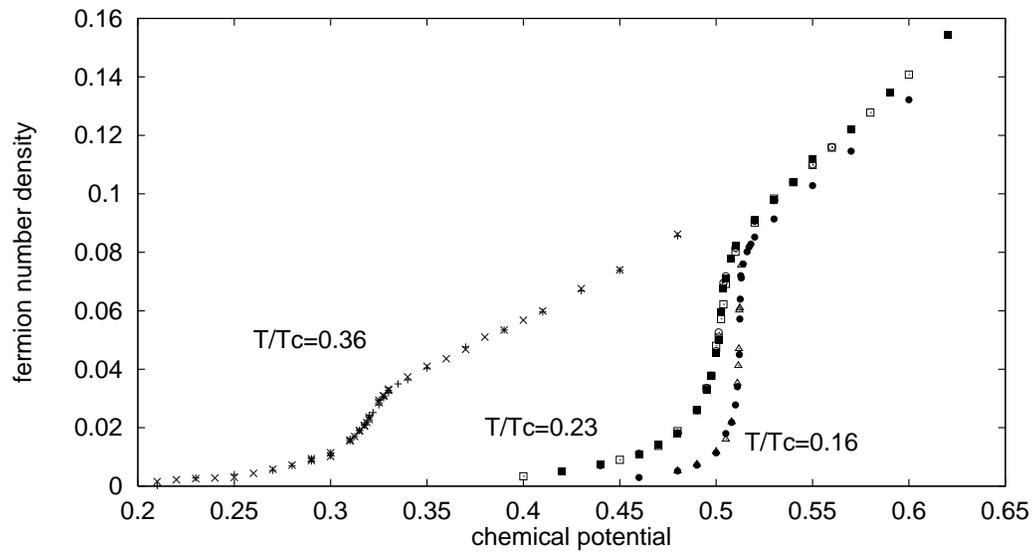}}

\smallskip
\caption[]{$n$ vs. $\mu$ for $T/T_c=0.36$, $(L_s=32, 48, 64)$; $T/T_c=0.23$  
$(L_s=32, 48, 64)$ and $T/T_c=0.16$ $(L_s=48, 72)$. The sizes of the error bars
are comparable to the sizes of the data points.}
\label{fig:density_mu.non0_t.non0}
\end{figure}

We first present the results for the order parameter $\Sigma$ and the fermion 
number density $n$ versus $\mu$ 
in Figs.~\ref{fig:sigma_mu.non0_t.non0} and
\ref{fig:density_mu.non0_t.non0} respectively. In each case we plot the values
of the observables for various values of the lattice spatial extend $L_s$.
For $T/T_c=0.36$, $L_s=32, 48, 64$; for $T/T_c=0.23$, $L_s=32, 48, 64$
and for $T/T_c=0.16$, $L_s=48, 72$.
These figures show a marked contrast with the corresponding ones 
at $T=0$, $\mu \neq 0$ discussed in the previous section.
In each case an asymptotic behavior is approached without mutual crossings of curves
for different $L_s$, in contrast to Figs.~\ref{fig:sigma_t=0_mu.nonzero} and 
\ref{fig:density_t=0_mu.nonzero}. However, as the temperature is lowered the variations of
the thermodynamic observables become sharper near the transition and it is therefore very 
difficult to distinguish between a weak first order and a second order transition.
It should also ne noted that in Fig.~\ref{fig:density_mu.non0_t.non0} the values of
the non-zero $n$ before the transition do not depend on $L_s$. This implies that 
the tail in $n$ for $\mu < \mu_c$ is not a finite $L_s$ effect; 
it is instead a finite $L_t$ or non-zero temperature effect.

\begin{figure}
  \begin{center}
    \begin{tabular}{c}
%      (a)  action cumulant minima\\
      \centerline{ \epsfysize= 3.0in
                   \epsfbox{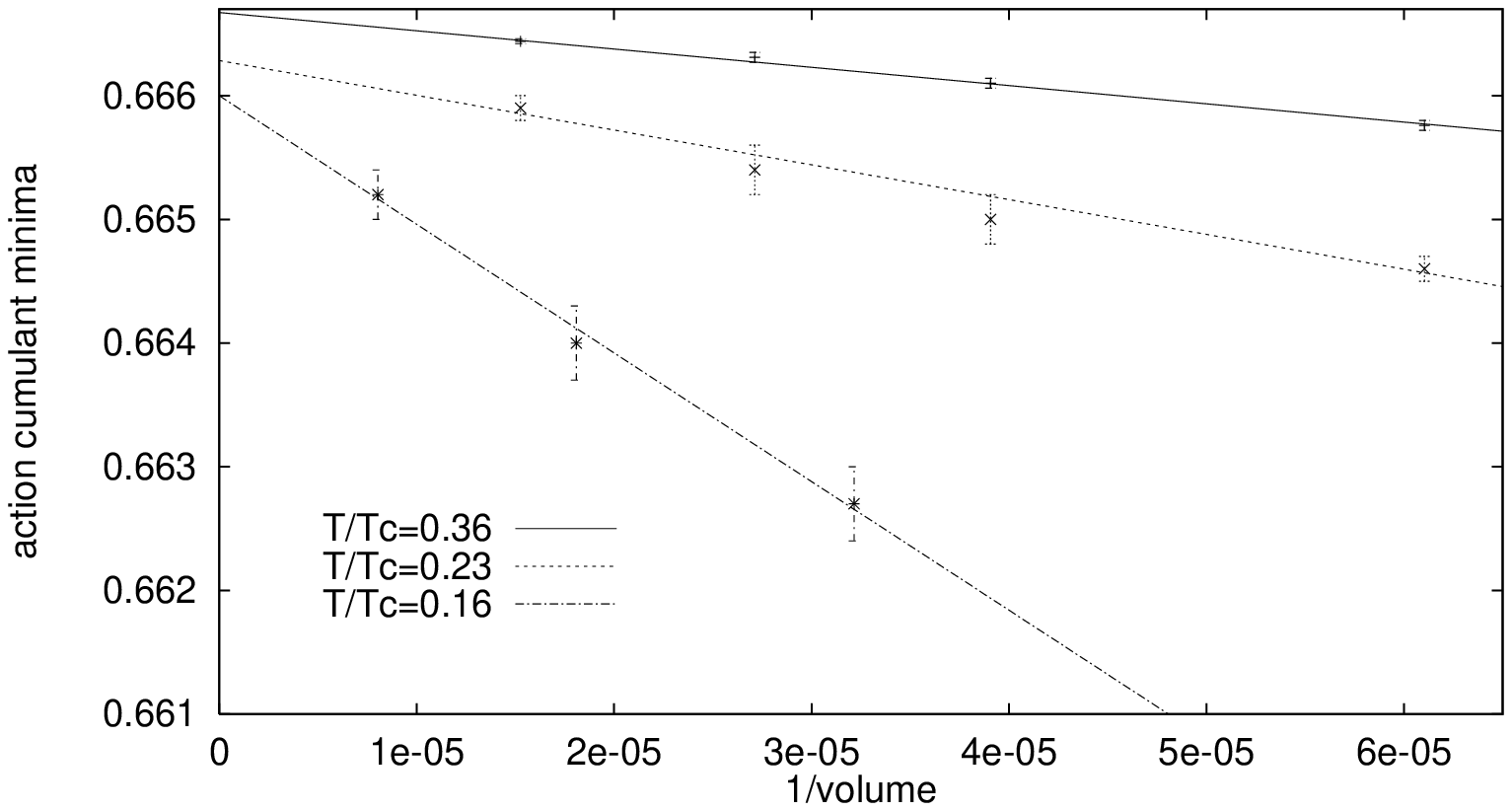}} \\ \\ \\
%      (b) energy cumulant minima \\
       \centerline{ \epsfysize= 3.0in
                   \epsfbox{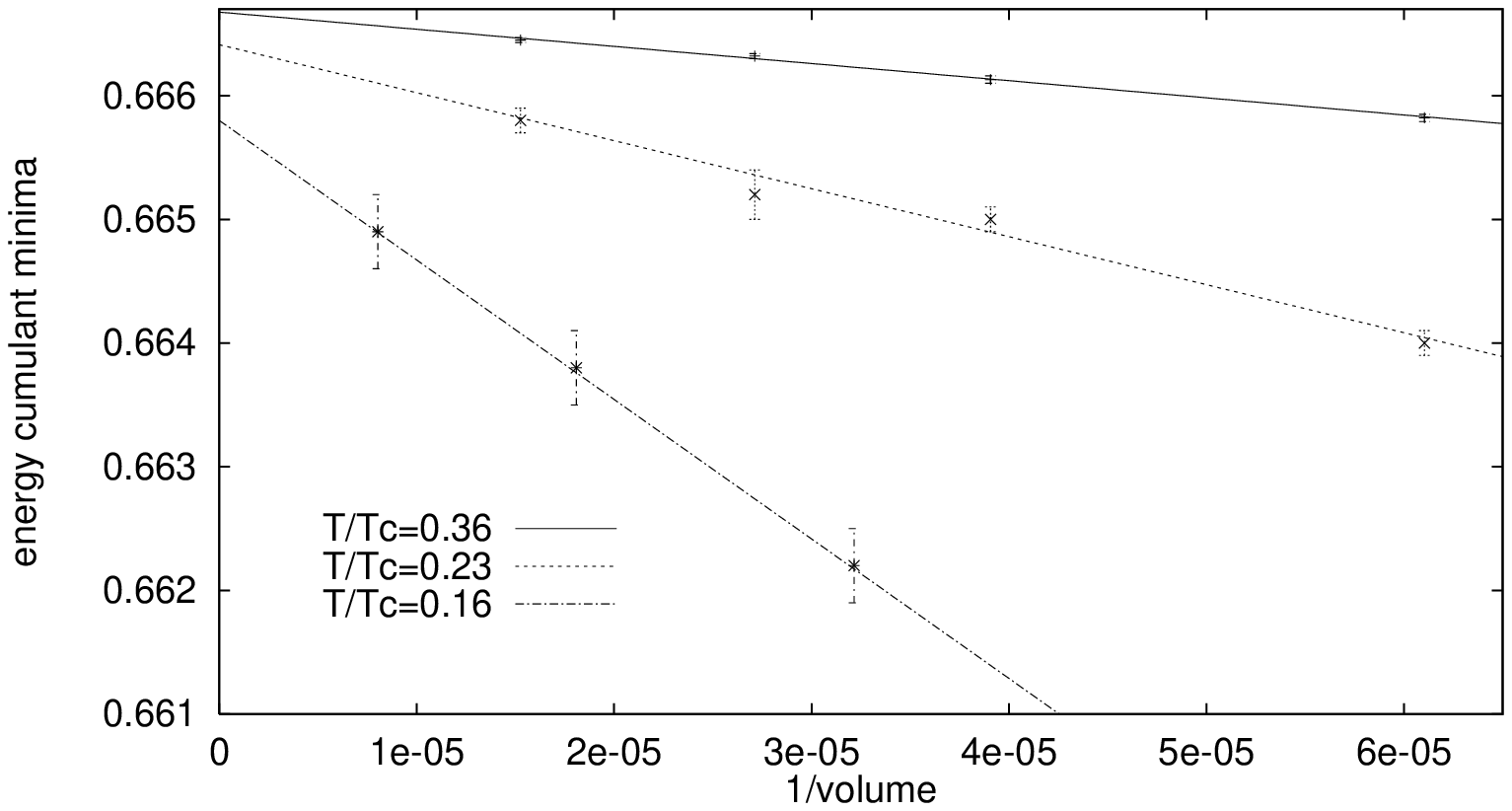}}
    \end{tabular}
  \end{center}
  \caption{The minimum values of the bosonic action and energy density cumulants vs. inverse
volume for $T/T_c=0.16, 0.23, 0.36$.
           The solid lines represent the best linear fits to the data.}
\label{fig:cum.all.min}
\end{figure}

For a more accurate determination of the order of the transition we performed 
again a more detailed finite size scaling analysis. 
We estimated the reduced cumulants $B_S$ for the bosonic action 
and $B_{E}$ for the internal energy density. 
For a precise calculation of these quantities we accumulated $30,000 - 50,000$
trajectories for the various lattices near $\mu_c$ at $T/T_c=0.36$
and $90,000 - 120,000$ trajectories at $T/T_c=0.23$ and $T/T_c=0.16$.  
We plot $B_{E}$ and $B_S$ versus $\mu$ in Figs.~[10-14]. 
Figure~\ref{fig:cum.all.min} shows the minimum values of $B_{E}$ and $B_S$ as functions
of the inverse volume. The straight lines represent the best linear fits to
the data.
Their $y$-intercepts for the three different parameter sets are :  
$B_S=0.66667(3)$, $B_{E}=0.66668(3)$ for $T/T_c=0.36$; $B_S=0.66628(13)$, 
$B_{E}=0.66641(13)$ 
for $T/T_c=0.23$
and $B_S=0.6660(3)$, $B_{E}=0.6658(4)$ for $T/T_c=0.16$. The results of this analysis confirm  
that the transition at $T/T_c=0.36$ is a second order transition. They also suggest
that at the two lower temperatures the transitions are weak first order, although
the possibility that these might also be second order transitions cannot be excluded
decisively. 
Studies on much larger lattices are required in order to detect the order of these
transitions conclusively. 
Nevertheless, the steep variations of the thermodynamic quantities near the transitions
at $T/T_c=0.23, 0.16$ together with the deviations of 
the thermodynamic limit of the cumulants' minima from the $\frac{2}{3}$ value support
the fact that the system is close to its tricritical point.

\section{Conclusions}
In this manuscript we presented the results of Monte Carlo simulations of the
three-dimensional
four-fermion model with $N=4$ fermion species at non-zero temperature
and density.
With the exception of confinement this model incorporates most of the
essential properties of QCD. The great advantage of this model is that
it is possible to perform lattice simulations at non-zero chemical
potential because its fermionic determinant remains real at $\mu \neq 0$.
We demonstrated using finite size scaling techniques  that the  $\mu
\neq 0$, $T=0$ phase transition is a first order transition. 
Our analysis did not succeed to distinguish the nuclear matter liquid-gas
transition from the chiral phase transition. Either the two transitions are
extremely close or the nuclear matter liquid-gas 
transition which is expected to occur at $\mu < \mu_c$ is very weak 
and is smoothed out by finite size effects or it doesn't exist in this model.

We have
also shown that the second order nature of the $T \neq 0$, $\mu=0$ phase
transition remains stable for  $T/T_c>0.23$ and $\mu/\mu_c<0.97$.
This imples that the large $N$ approximation (or mean field theory
description of the transition)  which predicts a second order phase
transition everywhere except at the isolated point $T=0$ 
\cite{hands_kocic_kogut.93}
describes the order of the transition on the largest part of the
critical line in the $(T, \mu)$ plane. In this sense
the model has a ``soft'' behavior on the biggest section of the critical
line. It should be noted
that the mean field approximation does not predict the
correct two-dimensional Ising universality
class \cite{kogut_98} of the second order phase transitions line. 
In that case non-perturbative effects are crucial in the critical 
region and the mean field results are modified by $1/N$ corrections.
However, the
results of this study show that corrections to the large $N$ limit for other
characteristics of the model, including the position of the
tricritical point and the position of the nuclear matter liquid-gas
transition, are very small indeed.

A simple continuity argument 
implies that the model has a tricritical point on the critical line
at a low temparature and very high density. Although our simulations
suggest that at low $T$ and high $\mu$ the phase
transition is a weak first order transition, studies on
much larger lattices would be required to ``locate'' the
tricritical point unambiguously.
Various approaches to QCD with two massless quarks at finite temperature
and density suggest the existence of a tricritical point on the boundary
of the phase with spontaneously broken chiral symmetry. The position of
the tricritical point in two-flavor QCD was estimated recently using a
random matrix model \cite{shuryak} and a Nambu$-$Jona-Lasinio model
\cite{berges} as $T_{tr} \approx 100$ MeV and $\mu_{tr} \approx 600-700$ MeV.
Possible experimental signatures suggested in \cite{stephanov99} should
allow future experiments at RHIC and LHC to  provide information about
the location and properties of this point. 

Another conclusion drawn from our work is that  lattices with very large
spatial extends are needed in order to observe
the order of the transition. Our studies on small volumes highlight the
difficulties that must be faced when trying to understand the critical
behavior of the thermodynamic limit by extrapolation from systems away
from this limit.  It was also shown that qualitative signatures of a
first order phase transition such as the variation of various
thermodynamic quantities near the transition  can be very misleading
unless they are accompanied by a detailed finite size scaling analysis.
This should be a warning to the lattice community that a quantitative 
understanding of the critical behavior of QCD at non-zero chemical 
potential lies along a difficult road.

\section*{Acknowledgements}
Discussions with Simon Hands, Misha Stephanov and Pavlos Vranas are greatly appreciated.
This work was supported in part by NSF grant PHY96-05199. 
CGS is supported by a Leverhulme Trust grant.
The computer simulations were done on the Cray C90's and J90's at NERSC
and on the NOW at SDSC.

\newpage

%\newpage

\newpage

\newpage

\newpage

\begin{figure}
  \begin{center}
    \begin{tabular}{c}
%      (a)  $L_s=32, 48$\\
      \centerline{ \epsfysize= 3.0in
                   \epsfbox{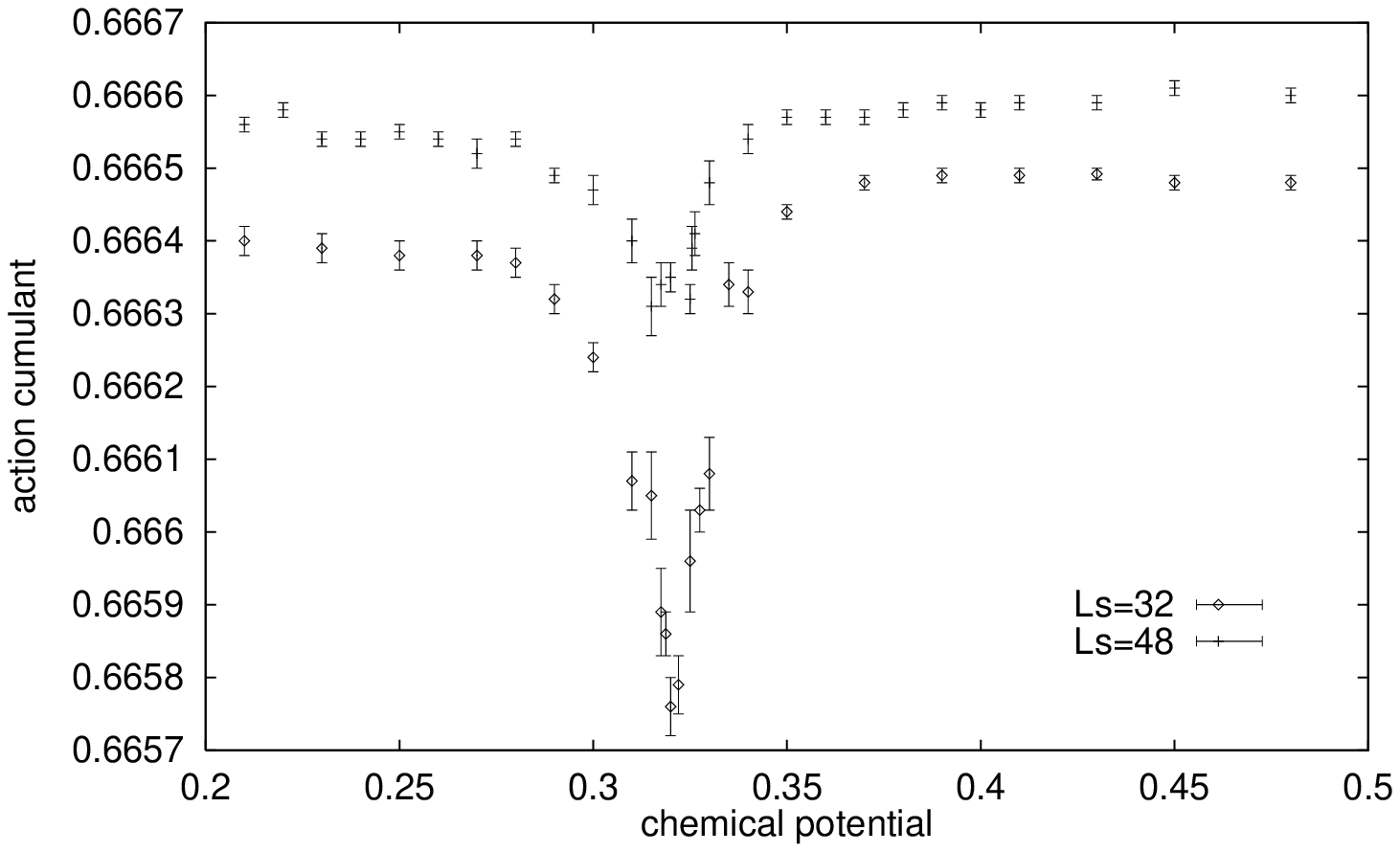}} \\ \\ \\
%      (b) $L_s=40, 64$ \\
       \centerline{ \epsfysize= 3.0in
                   \epsfbox{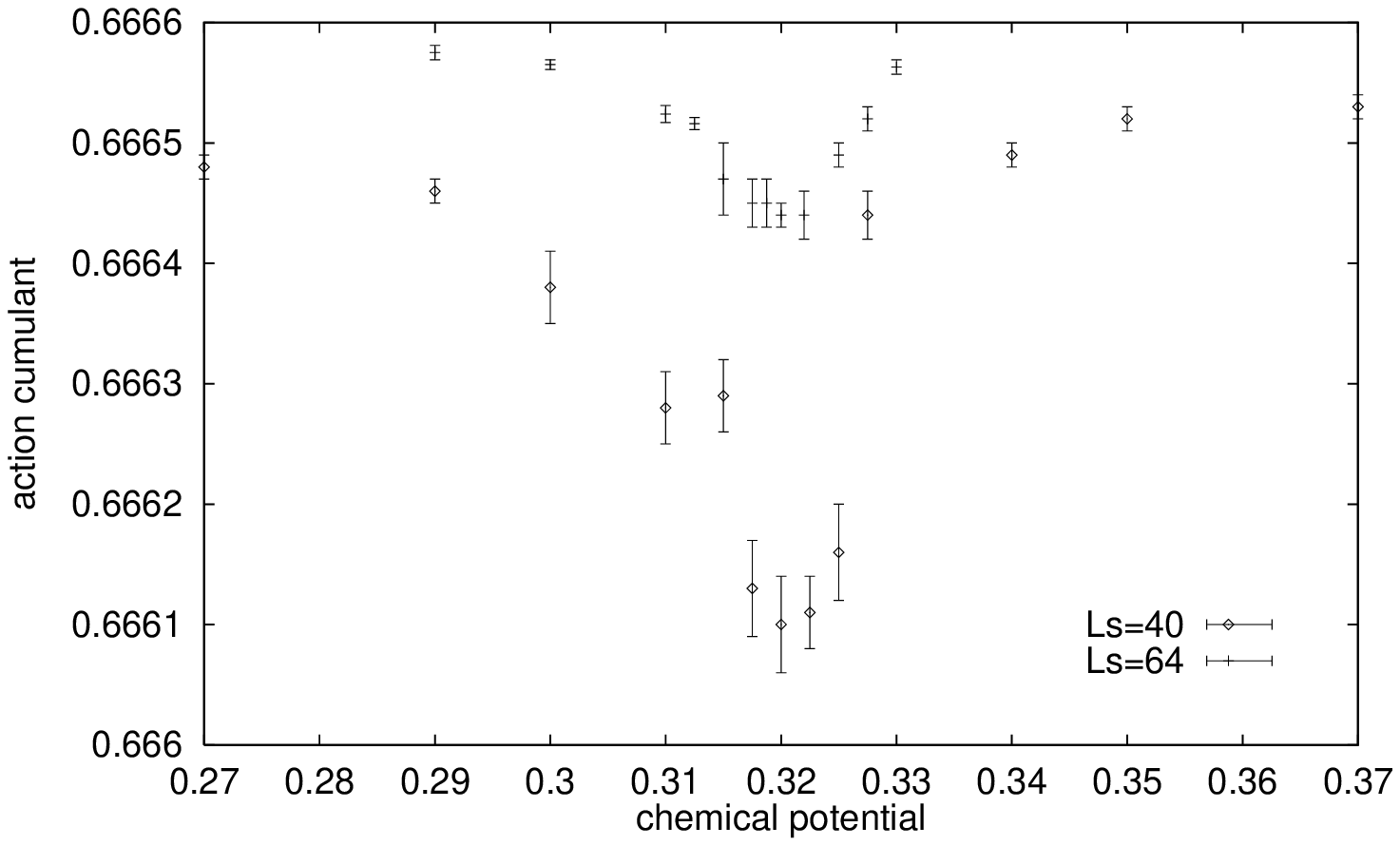}}
    \end{tabular}
  \end{center}
  \caption{The bosonic action cumulant $B_S$ vs. $\mu$ for $T/T_c=0.36$
and $L_s=32, 40, 48, 64$.} 
\label{fig:cumulant.s_t.nonzero_mu.nonzero}
\end{figure}

\begin{figure}
  \begin{center}
    \begin{tabular}{c}
%      (a)  $L_s=32, 48$\\
      \centerline{ \epsfysize= 3.0in
                   \epsfbox{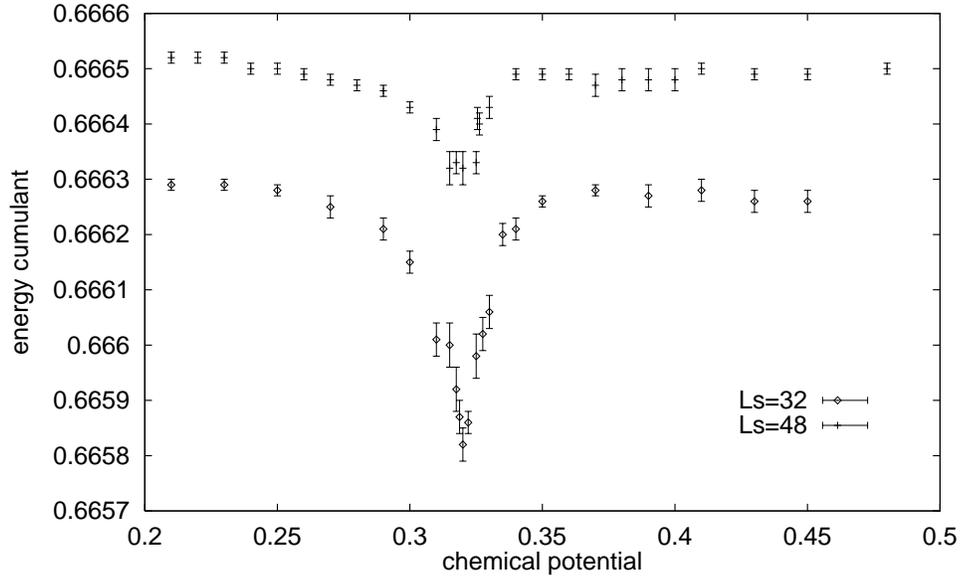}} \\ \\ \\
%      (b) $L_s=40, 64$ \\
       \centerline{ \epsfysize= 3.0in
                   \epsfbox{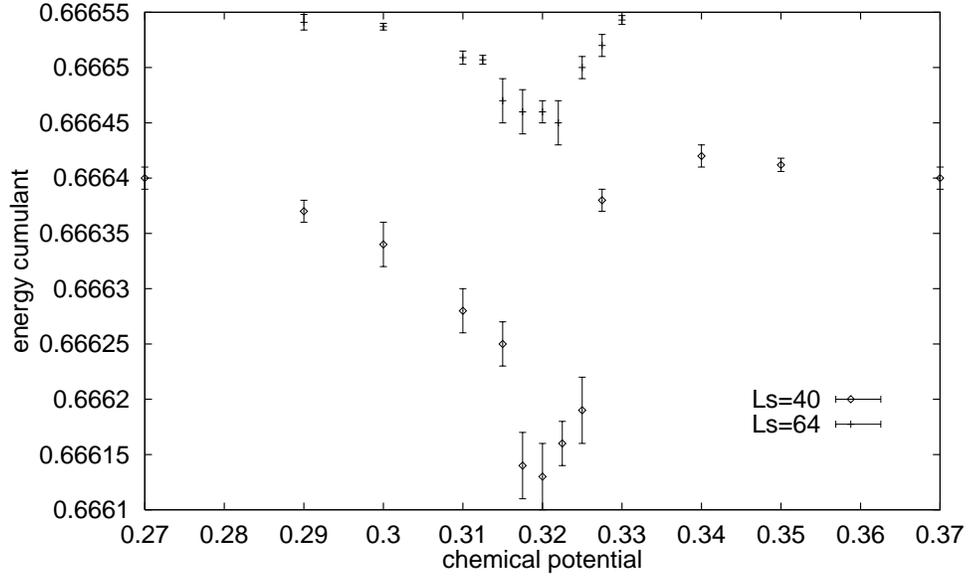}}
    \end{tabular}
  \end{center}
  \caption{The energy density cumulant $B_{E}$ vs. $\mu$ for $T/T_c=0.36$ and 
$L_s=32, 40, 48, 64$.}
\label{fig:cumulant.e_t.nonzero_mu.nonzero}
\end{figure}

\begin{figure}
  \begin{center}
    \begin{tabular}{c}
%      (a)  $L_s=32, 48$\\
      \centerline{ \epsfysize= 3.0in
                   \epsfbox{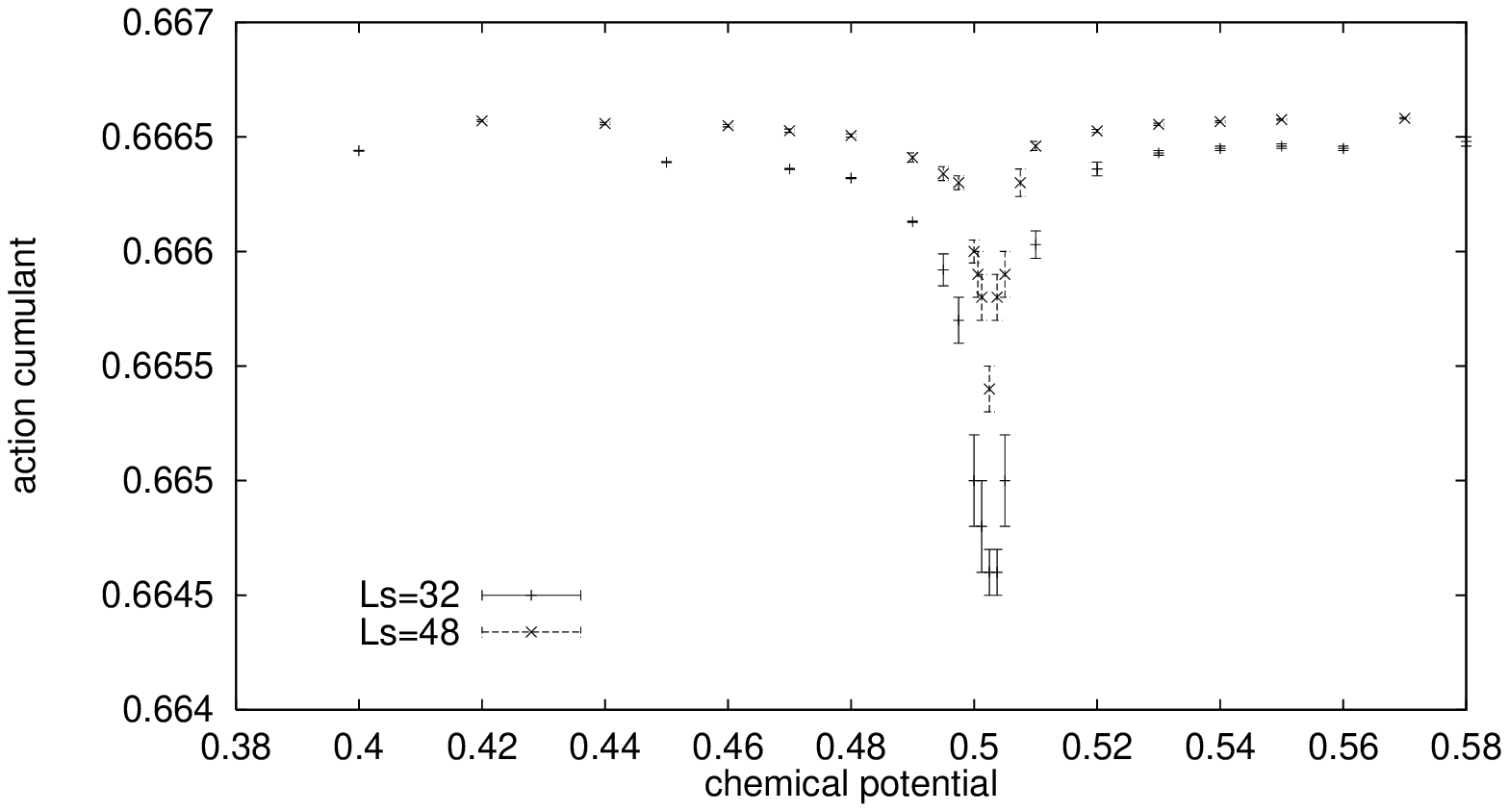}} \\ \\ \\
%      (b) $L_s=40, 64$ \\
       \centerline{ \epsfysize= 3.0in
                   \epsfbox{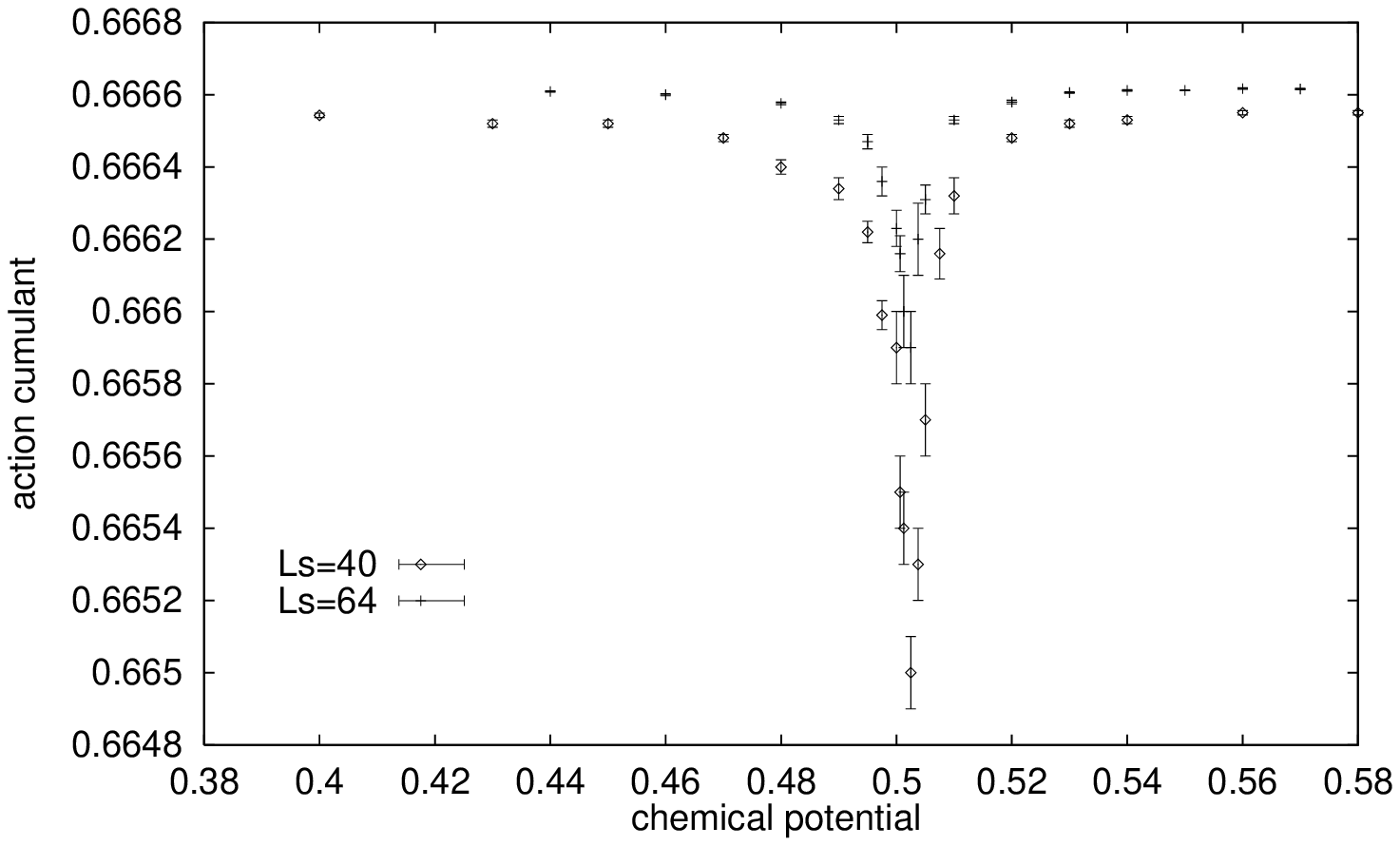}}
    \end{tabular}
  \end{center}
  \caption{The bosonic action cumulant $B_S$ vs. $\mu$ for $T/T_c=0.23$
 and $L_s=32, 40, 48, 64$.} 
\label{fig:cum.s.b55}
\end{figure}

\begin{figure}
  \begin{center}
    \begin{tabular}{c}
%     (a)  $L_s=32, 48$\\
      \centerline{ \epsfysize= 3.0in
                   \epsfbox{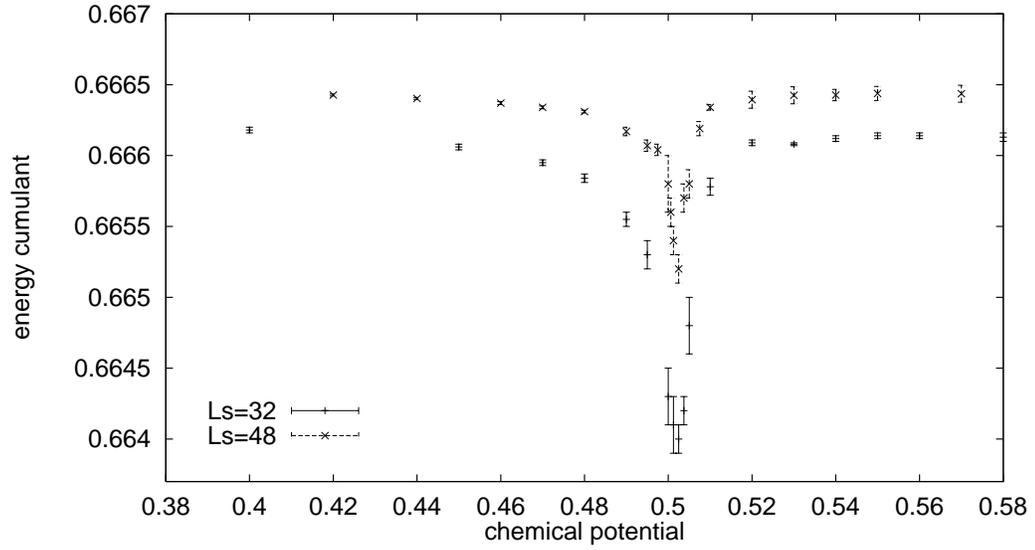}} \\ \\ \\
%      (b) $L_s=40, 64$ \\
       \centerline{ \epsfysize= 3.0in
                   \epsfbox{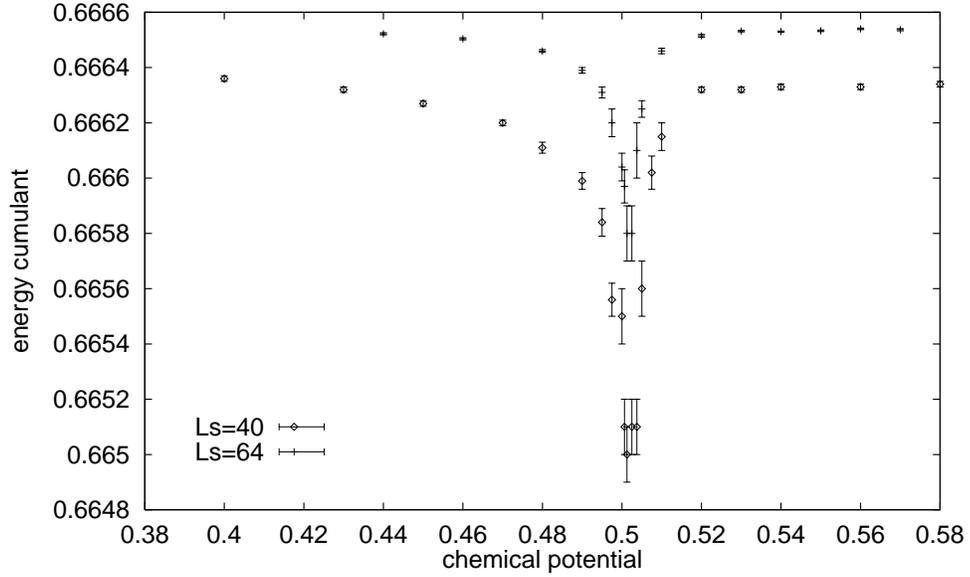}}
    \end{tabular}
  \end{center}
  \caption{The internal energy cumulant $B_{E}$ vs. $\mu$ 
           for $T/T_c=0.23$ and $L_s=32, 40, 48, 64$.}
\label{fig:cum.e.b55}
\end{figure}

\begin{figure}
  \begin{center}
    \begin{tabular}{c}
      \centerline{ \epsfysize= 3.0in
                   \epsfbox{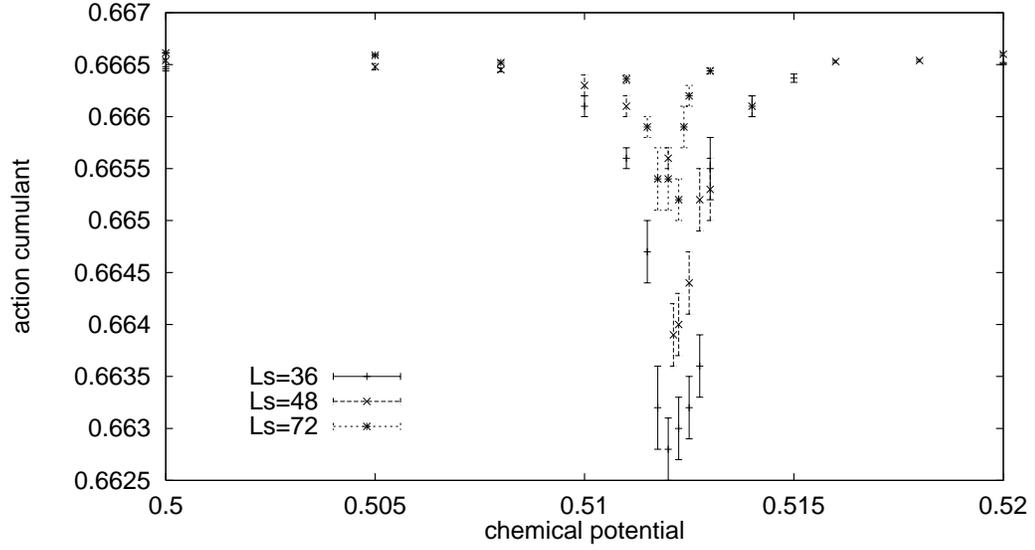}} \\ \\ \\
%      (b) $L_s=40, 64$ \\
       \centerline{ \epsfysize= 3.0in
                   \epsfbox{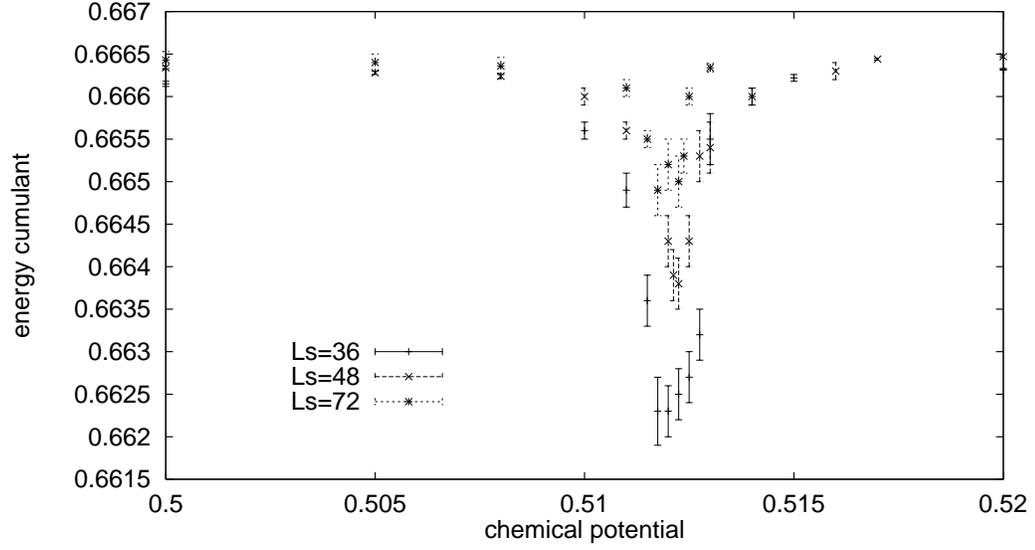}}
    \end{tabular}
  \end{center}
  \caption{The bosonic action and energy density cumulants $B_{E}$ and $B_S$ vs. $\mu$
           for $T/T_c=0.16$ and $L_s=36, 48, 72$.}
\label{fig:cum.e.b55}
\end{figure}

\end{document}